\documentclass[journal,twoside,web]{ieeecolor}
\usepackage{generic}
\usepackage{cite}
\usepackage{amsmath,amssymb,amsfonts}
\usepackage{algorithmic}
\usepackage{graphicx}
\usepackage{algorithm,algorithmic}
\usepackage{hyperref}
\usepackage{booktabs}
\hypersetup{hidelinks=true}
\usepackage{textcomp}
\def\BibTeX{{\rm B\kern-.05em{\sc i\kern-.025em b}\kern-.08em
    T\kern-.1667em\lower.7ex\hbox{E}\kern-.125emX}}
\begin{document}
\title{A Sonomyography-based Muscle Computer Interface for Individuals with Spinal Cord Injury}
\author{Manikandan Shenbagam, Anne Tryphosa Kamatham,\IEEEmembership{Student Member, IEEE}, Priyanka Vijay, Suman Salimath, Shriniwas Patwardhan, Siddhartha Sikdar, Chitra Kataria and Biswarup Mukherjee \IEEEmembership{Senior Member, IEEE}
\thanks{This manuscript was compiled on \today. This work has been partially funded by the Science and Engineering Research Board, Department of Science and Technology, Government of India, through a Core Research Grant (CRG/2021/004967, PI: BM). }
\thanks{Manikandan Shenbagam, Anne Tryphosa Kamatham are with the Centre for Biomedical Engineering, Indian Institute of Technology Delhi, New Delhi, Delhi, 110016. }
\thanks{Shriniwas Patwardhan and Siddhartha Sikdar are with the Department of Bioengineering, George Mason University, Fairfax, VA, 220230.}
\thanks{Priyanka Vijay, Suman Salimath, and Chitra Kataria are with 
the Institute of Rehabilitation Sciences, Indian Spinal Injuries Centre, New Delhi, Delhi, 110070.}
\thanks{Biswarup Mukherjee is with 
the Centre for Biomedical Engineering, Indian Institute of Technology Delhi and with the Department of Biomedical Engineering, All India Institute of Medical Sciences, New Delhi, 110029 (e-mail: bmukherjee@cbme.iitd.ac.in).}}
\maketitle

\begin{abstract}
Impairment of hand functions in individuals with spinal cord injury (SCI) severely disrupts activities of daily living. Recent advances have enabled rehabilitation assisted by robotic devices to augment the residual function of the muscles. Traditionally, non-invasive electromyography-based peripheral neural interfaces have been utilized to sense volitional motor intent to drive robotic assistive devices. However, the dexterity and fidelity of control that can be achieved with electromyography-based control have been limited due to inherent limitations in signal quality. We have developed and tested a muscle-computer interface (MCI) utilizing sonomyography to provide control of a virtual cursor for individuals with motor-incomplete spinal cord injury. We demonstrate that individuals with SCI successfully gained control of a virtual cursor by utilizing contractions of muscles of the wrist joint. The sonomyography-based interface enabled control of the cursor at multiple graded levels demonstrating the ability to achieve accurate and stable endpoint control. Our sonomyography-based muscle-computer interface can enable dexterous control of upper-extremity assistive devices for individuals with motor-incomplete SCI.
\end{abstract}


\section{Introduction}
\label{sec:introduction}
\IEEEPARstart{S}{pinal} cord injuries (SCI) disrupt the normal functioning of the spinal cord at the distal level of the injury. An incomplete injury results in the partial preservation of sensory and motor functions below the neurological level, including the lower sacral segments. The impairment of hand function frequently arises from traumatic cervical SCI, leading to difficulties in tasks involving grasping and manipulation due to either partial or complete paralysis of the hand muscles. Consequently, individuals with such impairments often experience reduced independence in daily activities and limited participation in recreational pursuits. Therefore, the need for improvement in hand function continues to be a priority for individuals with SCI~\cite{Snoek2004, Anderson2004}. 

Exercise therapy is the clinical standard for the rehabilitation of motor function ~\cite{Kloosterman2009, Lu2015}. However, several studies have evaluated key technologies such as transcranial magnetic stimulation~\cite{GomesOsman2015} and functional electrical stimulation~\cite{Popovic2006, Beekhuizen2008}, to supplement the benefits of exercise therapy but with limited success. Recently, advances in robotics have led to exciting new advances in the traditional rehabilitation protocols and standard of care~\cite{ Malley2006, Krebs1998}. Early attempts such as MIT-MANUS~\cite{Krebs1998} and MIME~\cite{Burgar2000, Lum2002} were restricted to proximal joints of the upper extremity. Later exoskeletons and assistive devices were developed to enable assisted movement of distal joints of the upper extremity, such as the MAHI Exoskeleton~\cite{Gupta2006, Sledd2006} and RiceWrist~\cite{Gupta2008}. Robots capable of multiple degrees-of-freedom simultaneous actuation of wrist, shoulder, and elbow joints have also been reported~\cite{Sugar2007, Perry2007}. The efficacy of the rehabilitation therapy is strongly linked to the methods employed to control the robotic end-effector. Two strategies have been widely reported: 1) passive control strategy and 2) cooperative or active control strategy. The passive strategy is used for individuals with complete loss of motor function to provide repetitive training of reaching and grasping activities along predetermined trajectories without any involvement of voluntary movement from the patient~\cite{ Carignan2008, Jiang2012, Li2015}. However, in active strategies, the user’s voluntary motor intent is sensed, and the residual motion is augmented and assisted by the exoskeleton. Maximizing the use of residual motor function promotes neuroplasticity in addition to improving muscle function by preventing disuse-induced muscle atrophy~\cite{Singh2021, Hogan2006}. 

Typically, motor intent is sensed by surface electromyography (EMG) electrodes on the affected limb~\cite{Zhang2013}. EMG-based methods that utilize biomechanical models to estimate the joint kinetics have been proposed~\cite{Fleischer2008, Rosen2001, Cavallaro2006, Noda2013}. Artificial neural networks and statistical learning methods~\cite{ Liu2013, McDonald2020, Li2023, Waris2019}  have also been proposed to decode and classify the patterns within the EMG signals. Proportional control of the joints has been achieved by computing simple EMG-derived temporal features~\cite{Ferris2006, Lenzi2012}. However, pattern-recognition based grasp decoding algorithms lack the ability to obtain graded proportional control due to inherently poor amplitude resolution and low signal-to-noise ratio of EMG signals, especially those obtained with dry electrodes~\cite{Clancy2002, Daley2012}.

In recent years, sonomyography, or ultrasound-based sensing of mechanical muscle contractions, has emerged as a potential alternative to electromyography for muscle activity monitoring and has been utilized for prosthetic control ~\cite{Jahanandish2019}. Ultrasound imaging is a non-invasive sensing modality with the ability to resolve deep-seated muscle compartments ~\cite{Ling2013, Li2014}. Sonomyography has been shown to be useful for detecting individual finger positions~\cite{ Castellini2012, Sikdar2014}, joint forces, and has been utilized to achieve proportional control of multiple degrees of freedom in order to control multi-articulated prosthetic devices~\cite{Sikdar2014, Akhlaghi2016, Dhawan2019}. Sonomyography signals are derived from physical muscle contractions as opposed to the motor unit action potentials in EMG. Therefore, sonomyography-based positional control relies on the extent of physical muscle deformation produced by the user and is directly congruent to natural motor control strategies employed by healthy individuals ~\cite{ Dhawan2019}. In this study, we develop a sonomyography-based muscle computer interface for individuals with motor-incomplete cervical spine injuries. The system is used to perform a virtual target achievement task to quantify and compare the performance of individuals with SCI to healthy controls. 
\section{Materials and methods}

\label{sec:MatMethods}

\subsection{Subjects}
We recruited nine individuals with motor-incomplete SCI (Mean age: 29$\pm$12) and nine able-bodied subjects (Mean age: 27$\pm$3) for this study. Written informed consent was obtained from all subjects. Ethical approval for conducting the study was obtained from the respective Institute Ethics Committees (IEC) at Indian Spinal Injuries Centre, New Delhi, and at the Indian Institute of Technology, New Delhi (ISIC/RP/2022/05 and P021/P050, respectively).

Individuals with incomplete cervical spinal cord injury at C5, C5-C6, and C6 levels were recruited from the ISIC. The subjects demonstrated partial upper-extremity motor control with at least a score of 3 or higher on the manual muscle testing scale (MMT)~\cite{Michelle2007}. Subjects were recruited in the sub-acute, acute and chronic states with American Spinal Injury Association Impairment Scale (AIS) grades A–C~\cite{Roberts2017}.

Functional assessments of the individuals with SCI were performed using Jebsen Taylor Hand Function Test (JTHFT) for the dominant hand ~\cite{Sears2010}. Additionally, patients completed the capabilities of the upper extremity test (CUE-T) prior to performing the experiment~\cite{Marino2012}. Individuals with diabetes, chronic pain, neuropathies, muscle contractures, and cognitive or neurological deficits were excluded from the study. Demographic information and clinical assessment scores for all subjects have been detailed in Table\,\ref{tab: demographics}.

\begin{table}[ht]
    \caption{Demographics and clinical assessment score of individuals with spinal cord injury}    
    \centering  
           
    \begin{tabular}{ p{0.7cm} p{0.5cm} p{1cm} p{0.5cm}  p{1cm} p{1cm} p{1cm}}
    \toprule
    \textbf{Subject ID} & \textbf{Age} & \textbf{Time since injury (weeks)} & \textbf{ASIA Scale} &  \textbf{Injury level}  & \textbf{CUE score}   &\textbf{JTFHT Score (Dominant Hand)}    
    \\
    \midrule
    
    SCI1    & 16  & 21  & C        & C6           & 183          & 176.24  \\
    \midrule
    SCI2    & 28  & 52  & B        & C6           & 157          & 150.21  \\
   \midrule
    SCI3    & 33  & 730 & B        & C6           & 170          & 89.93   \\
    \midrule
    SCI4    & 55  & 3   & C        & C6           & 72           & 210.52  \\
    \midrule
    SCI5    & 24  & 16  & A        & C6           & 97           & 166.24  \\
    \midrule
    SCI6    & 22  & 12  & A        & C6           & 62           & 162.23  \\
    \midrule
    SCI7    & 22  & 130 & A        & C5 - C6      & 115          & 123.43  \\
    \midrule
    SCI8    & 26  & 130 & A        & C5           & 123          & 199.50  \\
    \midrule
    SCI9    & 29  & 32  & A        & C6           & 139          & 89.23   \\
   
    \bottomrule
    \multicolumn{5}{l}{\textsuperscript{*} All the subjects are male}\\
    \bottomrule\\
    
      \end{tabular}
    \label{tab:demographics}
    
\end{table}

\subsection{Experimental procedures}
\label{sec:exp_procedure}


\begin{figure*}[!ht]
    \centering
    \includegraphics[width = \textwidth]{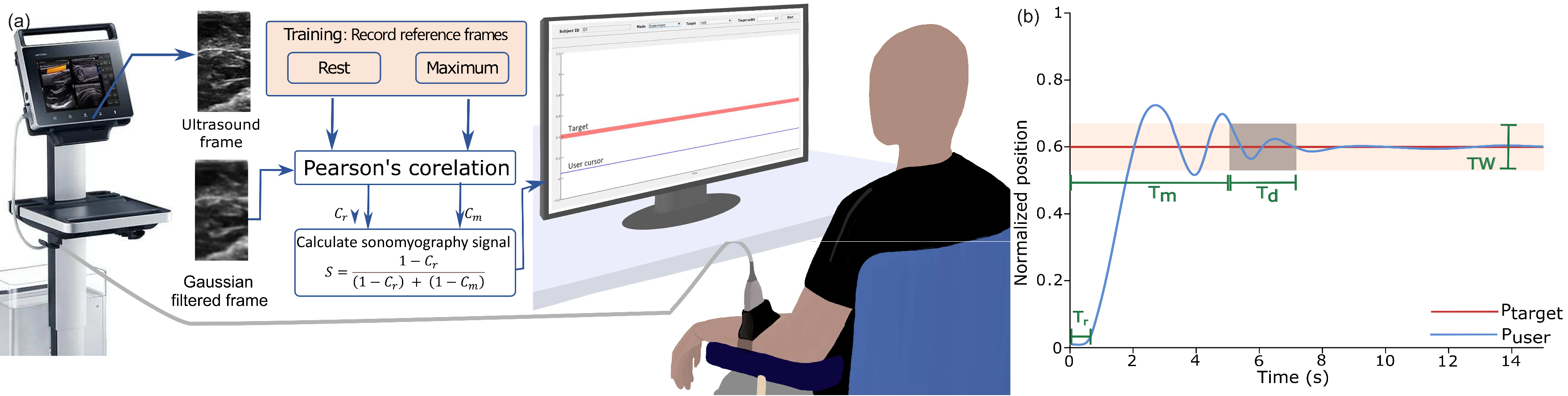}
    \caption{a. Experimental setup showing individuals with SCI instrumented with an ultrasound probe on the forearm. A clinical ultrasound machine captures the muscle activity as brightness-mode (B-mode) images. A MATLAB-based algorithm calculates the correlation between incoming and reference B-mode images and converts them to the cursor movement which is depicted on the screen.  
    b. Pictorial representation of a user cursor trajectory ($P_{user}$) and the target ($P_{target}$) showing the outcome metrics: movement time ($T_m$), target width($T_w$) and reaction time ($T_r$). The user was instructed to reach for the target of fixed width ($\pm$5\,\%) for a fixed amount of time defined by the dwell time $T_{d}$.}
    \label{fig:exp_setup}
\end{figure*}

Subjects were seated comfortably with their dominant arm placed on a platform such that the elbow rested directly below the shoulder on a padded armrest. A high-frequency linear ultrasound probe (Model: LV 7.5/60/128 Z-2, Samsung Electronics, South Korea) was connected to a clinical ultrasound imaging system (Model: PT60A, Samsung Electronics, South Korea) and was secured over the volar aspect of the forearm muscle using a custom-made 3D printed ultrasound probe holder using velcro straps as shown in Fig.~\ref{fig:exp_setup} (a). Ultrasound gel was applied between the probe and the skin. The probe was positioned transversely, aligning the field of view to image the major flexor muscles of the forearm. The imaging depth was set to 5\,cm to image the major flexor muscle groups of the forearm. Image sequences from the ultrasound system were streamed to a custom-developed MATLAB (Version 2022a, Mathworks Inc., USA) interface on a laptop PC (Model: Intel Core i7 – 7700HQ, 32GB RAM, 4GB NVIDIA GeForce GTX 1040Ti) in real-time using a USB-based video capture device (Model: Corsair Elgato Game Capture HD60S+, Corsair, USA). The approximate average frame rate obtained with the ultrasound acquisition was 20 frames per second.

All subjects were instructed via on-screen textual prompts and audio cues to perform a tenodesis grasping action by flexing the wrist joint for 30\,s. This was followed by 30\,s of rest, where the user was instructed to fully relax their wrist joint. The resulting ultrasound images of the maximum voluntary contraction corresponding to the fully flexed wrist position and the resting position were captured and stored in a training database. 

Computation of the sonomyography signal followed a procedure similar to that described by Dhawan et al.~\cite{Dhawan2019} and is summarised in section~\ref{sec:data_processing}. An on-screen cursor was displayed to the user where the vertical position of the cursor was proportionally mapped to the normalized sonomyography signal such that complete wrist flexion resulted in the cursor moving to the top of the screen (100\%) and relaxation of the wrist moved the cursor down to the bottom (0\%).

Based on the normalized value of two extrema of the tenodesis grasp, nine equispaced target positions (i.e., 10\,\%, 20\,\%, 30\,\%, 40\,\%, 50\,\%, 60\,\% 70\,\%, 80\,\% and 90\,\%) were presented to the subjects in random order. The width of the on-screen target was $\pm$5\,\%. Subjects were prompted by audio cues to reach the target and hold the cursor within the target continuously for 1.5\,s for the trial to be considered successful. Subjects were provided 10\,s to complete each target achievement task. After each trial, the participant was presented with a target at 0\,\% such that all new targets could be achieved from a resting wrist position. The cursor was not allowed to go beyond the 100\,\% level measured during calibration, and targets were restricted to 90\,\% in order to minimize fatigue and muscular injury. 

\subsection{Data processing}
\label{sec:data_processing}
The B-Mode ultrasound images of size 363x660 pixels were smoothed with a Gaussian filter with a two-dimensional kernel of 3x3  and standard deviation of 0.5 to remove speckle noise and minimize the effect of probe shifts~\cite{Dhawan2019}. Pearson's 2D correlation coefficient of the filtered incoming image frame was computed with the resting ($C_r$) and motion reference images ($C_m$) from the training database as shown in Fig.~\ref{fig:exp_setup}. The sonomyography signal ($S$) was computed as given in eq.\,\ref{eq:sonomyoraphy_signal}.

\begin{equation}
\label{eq:sonomyoraphy_signal}
    S = \frac{(1-C_m)}{(1-C_m) + (1-C_r)}
\end{equation}

The sonomyography signal ($S$) was normalized by utilizing upper and lower bounds obtained during the calibration step as described in section\,\ref{sec:exp_procedure}. However, it has been observed that sonomyography signals are prone to baseline drift due to probe shift, as well as large changes in user's posture~\cite{Kamatham2022-ICST, Dhawan2019}. We adopted a dynamic bound updation strategy developed by Dhawan et al. to mitigate the effects of the drift~\cite{Dhawan2019}. Briefly, this method updates the upper and lower normalization bounds of the sonomyography signal in real time by estimating the range of the signal dynamically. When the user attempts to exceed the current bounds, the bounds are expanded to accommodate the enhanced dynamic range. Similarly, the bounds are relaxed or shrunk if the user is unable to utilize the full range to match the dynamic range of the user's movement. 

\subsection{Outcome metrics}
The target achievement task was performed by able-bodied individuals and SCI subjects in a single session. The output trajectories were recorded during the task, and posthoc analysis was performed to compute the following outcome metrics:
   \subsubsection{Reaction time} The reaction time was measured as the time in seconds taken to initiate movement of the cursor (5\,\%) after the presentation of the target as shown in Fig.~\ref{fig:exp_setup}(b). The reaction time is an important measure of the information processing capacity of the neuromuscular system~\cite{Pachella1974}. 

    \subsubsection{Movement time} The movement time was defined as the time required by the user to reach the target from the resting position after the presentation of the target. The movement time is an important indicator of the efficiency of a human-machine interface. Fitt's law predicts that for any human-machine interaction task, there exists a linear relationship between the movement time and the index of difficulty of the task. The index of difficulty is the ratio of the distance of the target to its size or width expressed in a logarithmic scale (base 2)~\cite{Fitts1954}.

    \subsubsection{Success rate} A trial was considered successful only if the subject was able to reach and stay within the target bounds for a continuous period of 1.5\,s (see Fig.~\ref{fig:exp_setup}(b)). For each subject, the success rate was defined as the proportion of successful trials to the total number of trials for all target positions. The success rate  was calculated as,
        \begin{equation}
            Success\,rate = \frac{Number\ of\ successful\ targets}{Number\ of\ targets\ presented} \times 100 \%
        \end{equation}
    
    \subsubsection{Path efficiency} Path efficiency was computed as the ratio of the length of the path followed by the user to the path length of the ideal trajectory. The path length was computed as the cumulative Euclidean distance from the starting point of the trajectory to the point where the cursor first entered within the target bounds. This index provides a quantitative measure of the deviation of the user's trajectory from the ideal trajectory. 
        \begin{equation}
        \label{eq:path_eff}
            \eta = \frac{P_{target}}{\sum_{t=0}^{T_{m}}\lvert P_{user}(t+1) - P_{user} (t) \rvert} \times 100 \%
        \end{equation}
    \subsubsection{Endpoint error} Endpoint error ($E_{error}$) was computed as the mean absolute difference between the user's trajectory and intended target position after the acquisition of the target, i.e., within the 1.5\,s period when the cursor was within the target bounds (see Fig.~\ref{fig:exp_setup}(b)). Position error aims to provide a surrogate measure of positional error or bias during a stable grasping activity. It is calculated as,
        \begin{equation}
            E_{error} = \sum_{t=T_{m}}^{T_{m} + T_{d}} \frac{P_{target} (t) - P_{user} (t)}{N}
        \end{equation}
    Here, $N$ is the number of samples of $P_{user}(t)$, $T_{m}<t<T_{m} + T_{d}$, $P_{target}(t)$ is the position of the target presented to the user at time, $t$ and $P_{user} (t)$ is the position of the user's cursor at time, $t$.

    \subsubsection{Endpoint stability} The endpoint stability ($E_{stability}$ was calculated as the standard deviation of the cursor after target acquisition, i.e., within the 1.5\,s period when the cursor was within the target bounds. This metric aims to provide an indication of the jitter or instability of movement during grasping or holding activities. It is calculated as,

    \begin{equation}
        E_{stability} = \sum_{t=T_{m}}^{T_{m} + T_{d}} \frac{(P_{user} (t) - \mu_{user})^2}{N}
    \end{equation}
    Here, $N$ is the number of samples of $P_{user}(t)$, $t_{entry}<t<t_{entry} + t_{dwell}$ and, $\mu_{user}$ is defined as in Eq.~\ref{eq:mu_user},
    \begin{equation}
    \label{eq:mu_user}
        \mu_{user} = \sum_{t=T_{m}}^{T_{m} + T_{d}} \frac{P_{user} (t)}{N}
    \end{equation}

\subsubsection{Minimum jerk trajectory}
Flash et al. demonstrated through kinematic studies on human subjects performing unconstrained point-to-point movements that the movement trajectories can be computationally generated by minimizing the jerk of the movement~\cite{Flash1985}.  The objective function, $C$, to be minimized can be given as:

\begin{equation}
    C = \int_{0}^{T_m} \Big(\frac{d^3P}{dt^3}\Big)^2 \,dt
    \label{eq:MJT}
\end{equation}

Here, $P(t)$ is the position of the joint or cursor. The mathematical expressions for $P(t)$ which bring the minimization function in eq.~\ref{eq:MJT} to a minimum are given by eq.~\ref{eq:polynomial_MJT},

\begin{equation}
    P(t) =  P(0) + [P(0) - P(T_m)] [15\tau^4 - 6\tau^5 -10\tau^3]
    \label{eq:polynomial_MJT}
\end{equation}

\subsection{Statistical tests}

Due to the limited sample size and difficulty in testing normality assumptions, we performed non-parametric statistical tests. The Mann-Whitney U test was performed to compare the performance of able-bodied subjects and individuals with SCI for all outcome measures. The test was performed with the alternative that the distribution of success rate and path efficiency in the able-bodied group is shifted to the right of the distribution in the SCI group. For the position error, stability error, and movement time, the test was performed with the alternative hypothesis that the distribution in the able-bodied group is shifted to the left of the distribution in the SCI group. Further, Friedmann’s test was performed to determine the influence of target position on all the outcome metrics for both the able-bodied group as well as the SCI group~\cite{Hollander2013}. All statistical analyses were performed using IBM SPSS Statistics (Version 28.0, IBM Corp, Armonk, NY, USA).

\section{Results}

\begin{figure*}[!ht]
    \centering
    \includegraphics[width = \textwidth]{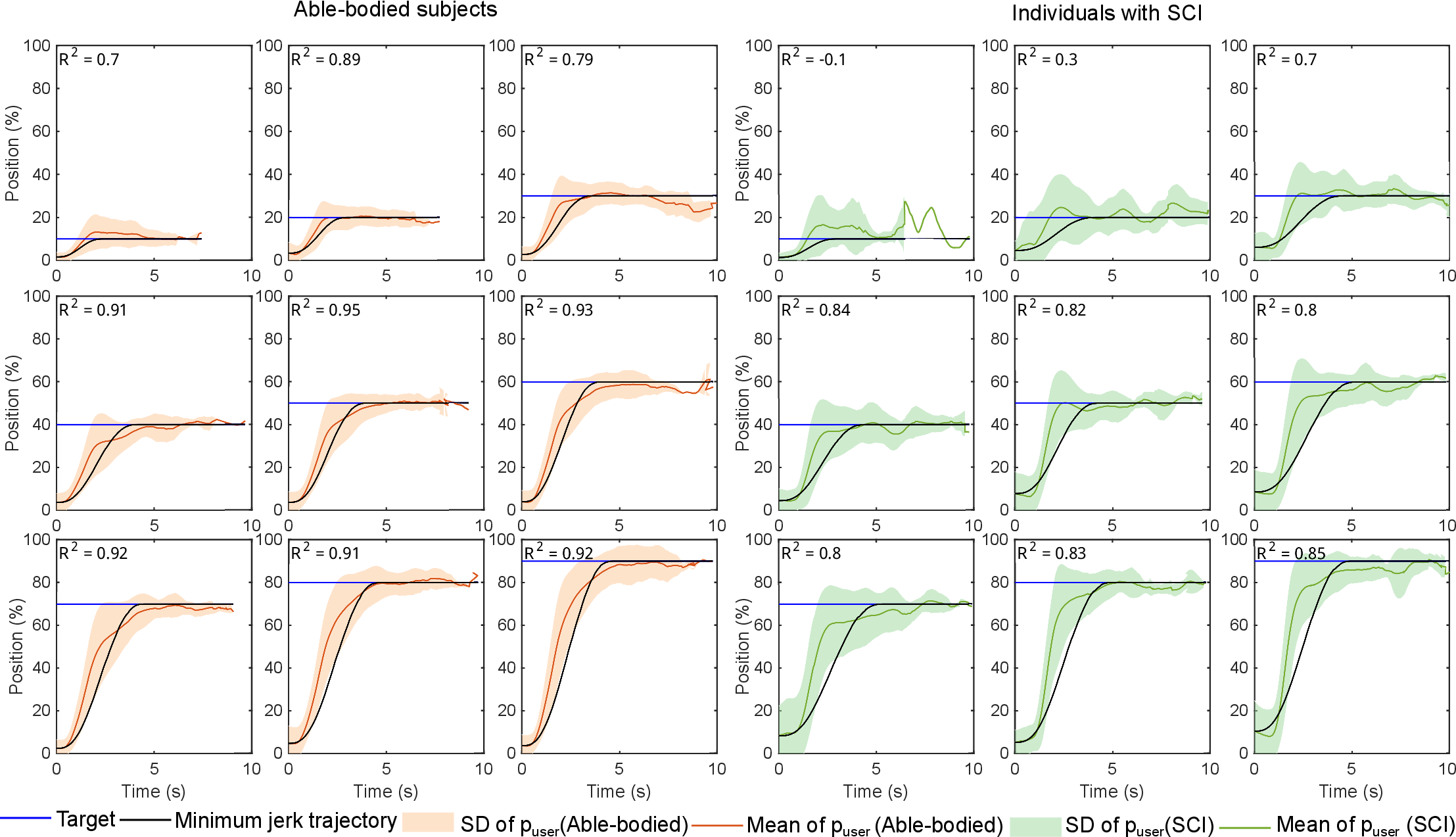}
    \caption{Time series plots showing mean user trajectory in the target achievement task for all target positions for able-bodied participants (orange) and individuals with SCI (green). The minimum jerk trajectory (black) shows the idealized path to attain the target. The coefficient of determination (R\textsuperscript{2}) provides a measure of the closeness of the observed user trajectory to the idealized trajectory.}
   \label{fig:postrajectory}
\end{figure*}

Supplementary movies S1, S2, and S3 demonstrate the process of user training, calibration, and testing phases for an individual with cervical spinal cord injury (see~\ref{sec:exp_procedure}). In the testing phase, targets appeared on the computer screen, and the user was prompted to attain the target using a cursor controlled by the sonomyography signal computed using the user's dynamic muscle activity. Fig.~\ref{fig:postrajectory} shows the temporal trajectories of the user-controlled cursor for all target positions for able-bodied subjects and individuals with SCI, respectively. 
It has been shown in several key studies that the joint kinematics of able-bodied individuals follow typical idealized minimum jerk trajectories in a variety of tasks such as arm reaching~\cite{Flash1985, Morasso1981}, vertical arm movements~\cite{Atkeson1985}, catching~\cite{Fligge2012}, and other motor tasks. We compared the sonomyography-dervied target acquisition trajectories produced by able-bodied subjects and individuals with SCI with the idealized minimum jerk trajectory as shown in Fig.~\ref{fig:postrajectory}. We hypothesized that the sonomyography-derived trajectories would closely follow the minimum jerk trajectory in able-bodied subjects but deviate from ideal behavior for individuals with SCI due to impaired neuromuscular function. Our results demonstrate that trajectories produced by able-bodied subjects were well correlated with the idealized trajectories across all target positions (R\textsuperscript{2} = 0.88$\pm$0.08). SCI subjects deviated from the ideal minimum jerk trajectory significantly (R\textsuperscript{2} = 0.67$\pm$0.28) compared to able-bodied subjects (p= 0.013, $U$ = 12.5, $Z$ = -2.478) confirming our hypothesis. 
\begin{figure} [!bh]
    \centering
    \includegraphics[width = \columnwidth]{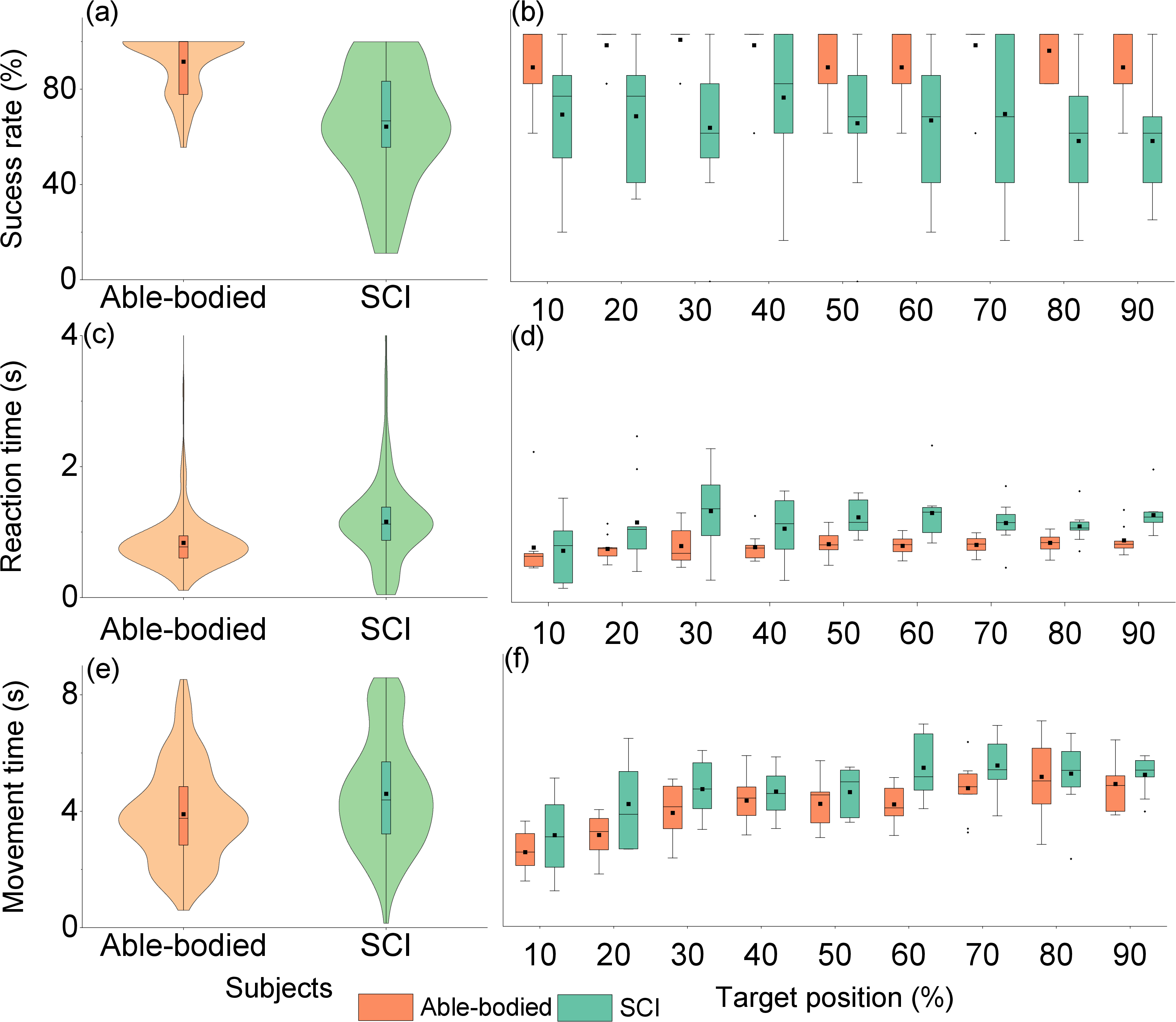}
    \caption{Outcome metrics derived from the temporal user trajectories showing the influence of the target position on (a), (b) success rate, (c), (d) reaction time, and (e), (f) movement time. Violin plots provide an overall comparison between able-bodied (orange) and individuals with SCI (green) for the above-mentioned outcome metrics. Box plots demonstrate the influence of the target position on the outcome metrics.}
    \label{fig:success_rate}
\end{figure}

Fig.~\ref{fig:success_rate}(a) shows the success rate for target acquisition for able-bodied subjects and individuals with SCI. As expected, our results demonstrate a statistically significant (p$<$0.001, $U$ = 1342, $Z$ = -7.16)  difference in the success rate between able-bodied subjects and individuals with SCI. The mean success rate for individuals with SCI was found to be 64.35$\pm$24.03\,\%, while able-bodied subjects demonstrated a success rate of 91.60$\pm$12.34\,\%. Fig.~\ref{fig:success_rate}(b) shows the effect of target position on the success rate for able-bodied subjects and individuals with SCI. We also found a statistically significant effect of target position on the success rate for individuals with SCI (p$<$0.001, \(\chi^2(8) = 29.511\)), but no effect was found for able-bodied subjects (p=0.236, \(\chi^2(8) = 10.431\)). 
\begin{figure} [!bh]
    \centering
    \includegraphics[width = \columnwidth]{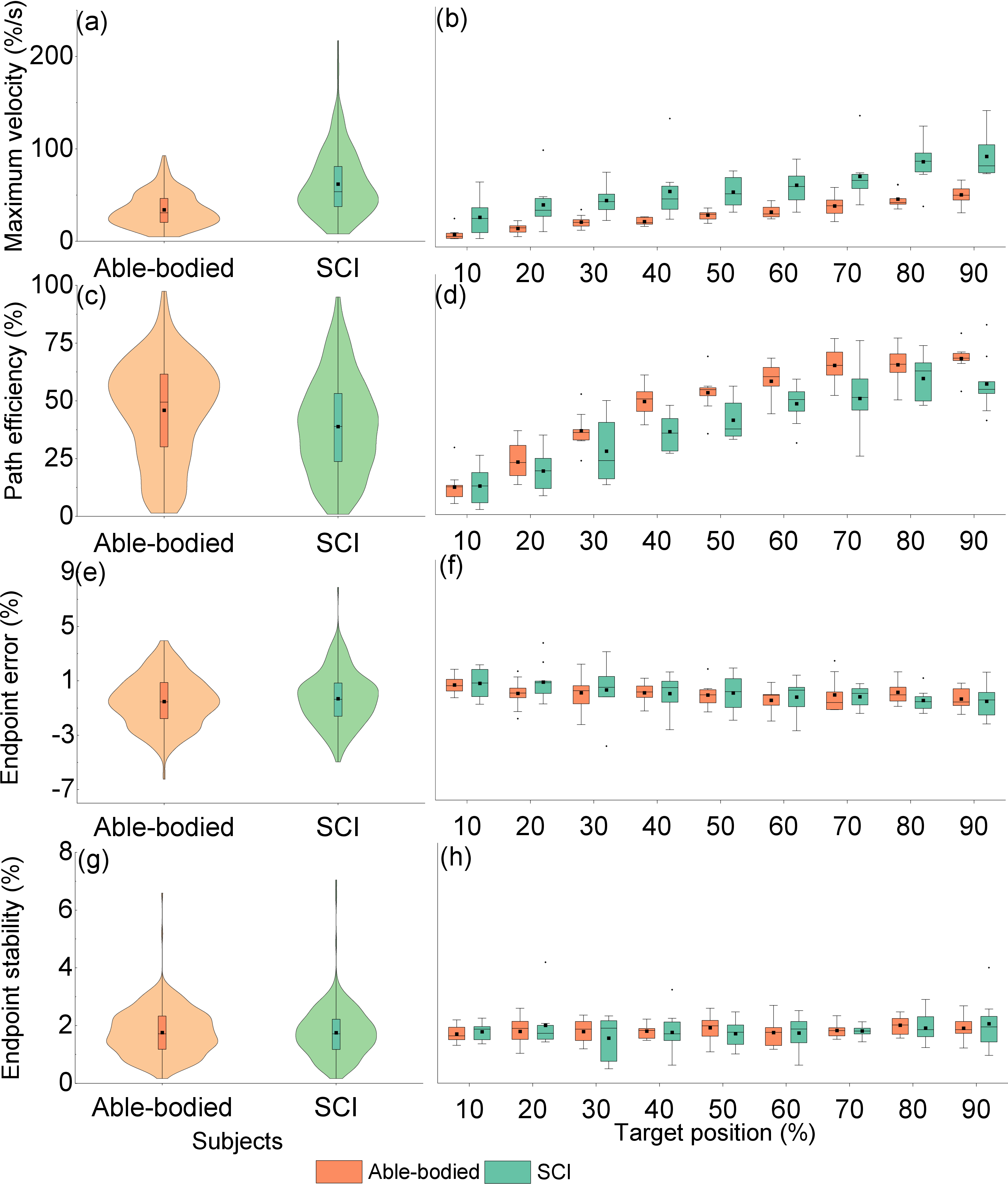}
    \caption{Outcome metrics derived from the temporal user trajectories showing the influence of the target position on (a), (b) maximum velocity, (c), (d) path efficiency, (e), (f) endpoint error and (g), (h) endpoint stability. Violin plots provide an overall comparison between able-bodied (orange) and individuals with SCI (green) for the above-mentioned outcome metrics. Box plots demonstrate the influence of the target position on the outcome metrics.}
    \label{fig:max_velocity}
\end{figure}
We computed the user's reaction time for each trial from the user's temporal trajectories as shown in Fig.~\ref{fig:success_rate}(a). There was a statistically significant difference (p$<$0.001, $U$ = 46719, $Z$ = -11.314) in the mean reaction time 1.16$\pm$0.64\,s for individuals with SCI compared to able-bodied individuals 0.84$\pm$0.45\,s  indicating that individuals with SCI are slower to initiate movements compared to able-bodied individuals. We found a significant effect of the target position on the reaction time for both groups (p$<$0.001, \(\chi^2(8) = 29.813\) for able-bodied subjects and p$<$0.001, \(\chi^2(8) = 33.956\) for individuals with SCI ). 

We also measured the time taken by the user to acquire the target from temporal trajectories of the user's cursor movement for able-bodied subjects and individuals with SCI (Fig~\ref{fig:success_rate}(c)). We found that individuals with SCI demonstrated higher mean movement times of 4.59$\pm$1.90\,s compared to able-bodied subjects of 3.90$\pm$1.62\,s. However, the difference in movement time was not statistically significant for the two groups (p=0.207, $U$ = 2867, $Z$ = -1.261). Fig~\ref{fig:success_rate}(d) shows the effect of target position on movement time. As expected, there was a statistically significant effect of target position on movement time in both able-bodied (p$<$0.001, \(\chi^2(8) = 46.993\)), as well as individuals with SCI (p$<$0.001,\(\chi^2(8) = 34.833\)), as predicted by Fitt's law~\cite{Fitts1954}. We also fit a linear regression model to estimate the throughput as described by Fitt's law. The throughput for individuals with SCI was found to be 1.23\,bits/s while that of able-bodied individuals was found to be 1.02\,bits/s ( 
see Fig.~S1).  

Fig~\ref{fig:max_velocity}(a) presents plots depicting the maximum velocity achieved by both able-bodied subjects and individuals with SCI. This metric represents the highest velocity achieved by the subjects in order to attain the on-screen target. The study revealed a statistically significant finding (p$<$0.001, $U$ = 23645, $Z$ = -11.781), indicating that individuals with SCI exhibited a notably higher maximum velocity of 61.88$\pm$34.52\,\%, in contrast to the velocity of 34.01$\pm$17.85\,\%  observed among the able-bodied subjects. Fig~\ref{fig:max_velocity}(b)  illustrates a plot showcasing the relationship between maximum velocity and the position of the target. Notably, there was a statistically significant impact of target position on the maximum velocity by both the able-bodied subjects (p$<$0.001, \(\chi^2(8) = 26.667\))  and individuals with SCI (p=0.003, \(\chi^2(8) = 23.147\)).

Fig~\ref{fig:max_velocity}(c) shows plots of path efficiency for able-bodied subjects and individuals with SCI. Path efficiency provides a comparative measure of the total path length of the user's temporal trajectory relative to the target position. This study found that individuals with SCI had statistically significantly lower (p$<$0.001, $U$ = 1323, $Z$ = -6.481) path efficiency of 38.89$\pm$20.82\,\% compared to able-bodied subjects of 45.92$\pm$21.45\,\%. Fig~\ref{fig:max_velocity}(d) shows a plot of path efficiency with respect to the target position. There was a statistically significant effect of target position on path efficiency in both able-bodied subjects (p$<$0.001, \(\chi^2(8) = 29.511\)) and individuals with SCI (p$<$0.001, \(\chi^2(8) = 29.00\)). 


We analyzed the stability and accuracy of the user's movement after attaining the target to better understand the endpoint error and jitter for able-bodied subjects and individuals with SCI. Figs.~\ref{fig:max_velocity}(e) and ~\ref{fig:max_velocity}(g) show that the mean endpoint error for individuals with SCI was found to be -0.30$\pm$1.94\,\% and -0.51$\pm$1.84\,\% for able-bodied subjects, while endpoint stability for able-bodied subjects and individuals with SCI was 1.76$\pm$0.79\,\% and for individuals with SCI was 1.75$\pm$0.86\,\% respectively. The difference in the endpoint stability (p=0.552, $U$ = 3064, $Z$ = -0.595) and error metrics (p=0.685, $U$ = 3120, $Z$ = -0.406) between the two groups were not statistically significant indicating that sonomyography enables stable control of the end-effector for individuals with SCI after attaining the target. Figs.~\ref{fig:max_velocity}(f) and~\ref{fig:max_velocity}(h) show that there is no significant influence of the target position on the endpoint error or endpoint stability for both able-bodied subjects (p$=$0.256, \(\chi^2(8) = 10.133\) and p$=$0.484, \(\chi^2(8) = 7.496\) respectively) and individuals with SCI (p$=$0.326, \(\chi^2(8) = 9.200\) and p$=$0.835, \(\chi^2(8) = 4.233\) respectively) providing stable endpoint control over the entire range of motion of the user. 

\section{Discussion}
In this study, we have developed and tested a sonomyography-based human-machine interface for individuals with motor-incomplete SCI to enable control of bionic devices. Individuals with C5, C5-C6, and C6 level motor incomplete cervical spine injury were recruited for our study. Age-matched able-bodied individuals were used as a control for our study. Participants were instructed to execute tenodesis grasping action with varying activation levels, while ultrasound images of the forearm muscles were obtained in real-time and processed to enable control of an on-screen virtual cursor. This task mimicked a muscle-machine interface and allowed for precise manipulation of the end-effector in response to muscle contractions.

\begin{figure*}[!ht]
    \centering
    \includegraphics[width = \textwidth]{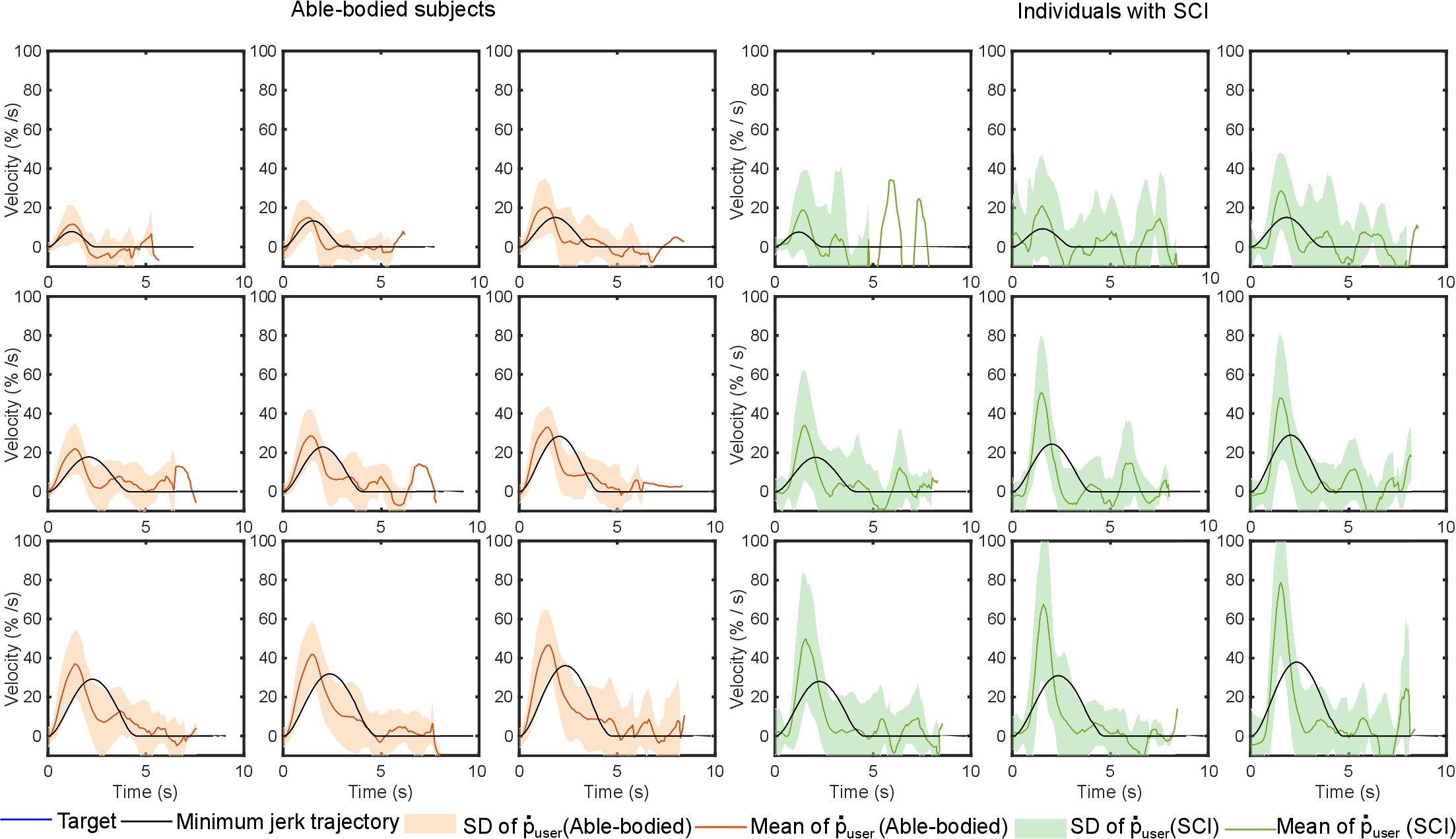}
    \caption{Time series plots showing the cursor velocity in the target achievement task for all target positions for able-bodied participants (orange) and individuals with SCI (green). The minimum jerk velocity trajectory (black) shows the idealized path to attain the target.}
   \label{fig:veltrajectory}
\end{figure*}

\subsection{Sonomyography detects muscle activity in individuals with motor incomplete SCI}
Our results demonstrate that sonomyography or ultrasound imaging is able to successfully detect muscle activity in individuals with SCI. Subjects recruited for our study had varying levels of residual hand function based on the level, severity, and time since injury, as demonstrated by their CUE and JTFHT scores. All subjects successfully attained approximately 64\,\% of the targets presented to them. SCI subjects experienced difficulty attaining the 90\,\% target position (Mean success rate at 90\,\% target level was 56.85\,\%). This could be attributed to the limited range of wrist joint motion for these subjects and fatigue experienced while attaining targets at the extrema of their range of motion. 

\subsection{Sonomyography enables stable endpoint control for individuals with SCI}

Results from this study show that the sonomyography-based muscle computer interface provides precise control of the cursor allowing the subjects to minimize targeting errors. Additionally, endpoint stability has also been shown to be similar for able-bodied subjects and individuals with SCI signifying that the user was able successfully to acquire and minimize the jitter of the end-effector. Hence, bionic devices utilizing sonomyography-based approaches to provide assistive functions may enable superior dexterity compared to surface electromyography-based techniques. Several studies have demonstrated that EMG-based pattern recognition systems require multiple sessions to familiarize and train the user to obtain stable classification outcomes~\cite{Powell2014, McDonald2020}. In contrast, SCI subjects were able to achieve online control of the sonomyography-based interface with no prior training.

\subsection{Individuals with SCI exhibit slower temporal dynamics compared to healthy controls with minimal training}
Analysis of the temporal dynamics of user-generated trajectories revealed that individuals with SCI have significantly higher reaction times compared to able-bodied individuals. This study also sheds light on crucial aspects of motor planning and execution of goal-oriented motor behavior for individuals with SCI. Studies have demonstrated that the motor response time when responding to a movement stimulus is extended in individuals with SCI in comparison to healthy controls~\cite{Labruyere2011, Labruyere2013}. This could potentially be linked to an impaired capacity to synchronize corticospinal descending volleys~\cite{Cirillo2016}.

Individuals with SCI also demonstrated significantly prolonged movement times across all target positions. The slowing of motor processes may be due to a decrease in wrist extensor muscle strength~\cite{Gronley2000, Wierzbicka1992}. Additionally, the slowing of motor outputs could also be a result of central adaptation to preserve endpoint accuracy and stability similar to healthy counterparts~\cite{Todorov2004}. This is supported by the observation that both endpoint error and endpoint stability were not significantly different in SCI participants compared to healthy controls. 

\subsection{Higher peak movement velocities for individuals with SCI}

Several studies have demonstrated that the endpoint velocity during goal-directed multi-joint as well as single-joint movements in able-bodied individuals follows a typical bell-shaped profile~\cite{Flash1985, Morasso1981, Atkeson1985}. However, tasks requiring precise endpoint control demonstrate velocity profiles that are characterized by several discrete sub-movements arising from a ballistic reaching phase and subsequent corrective phases to achieve precision during final target acquisition. It has also been demonstrated that the deceleration phase of the velocity profile is prolonged in healthy subjects as end-point accuracy requirements are made more stringent~\cite{Milner1990}. A close examination of the velocity trajectories for both able-bodied subjects and individuals with SCI shown in Fig.~\ref{fig:veltrajectory} reveals the presence of multiple discrete segments in addition to the primary bell-shaped profile.  Normalized velocity profiles for able-bodied subjects also show a pronounced secondary peak during the deceleration phase, indicating a controlled slowing of motor processes to improve aiming accuracy. The lengthened deceleration could be indicative of the engagement of the wrist extensor muscles to achieve the fine control required to successfully acquire the target. In contrast, a qualitative comparison of the velocity profile for individuals with SCI and the idealized minimum jerk velocity profile reveals a more pronounced primary velocity peak. Additionally, the primary movement velocity profile appears to have symmetric acceleration and deceleration phases. This may be attributed to the inability to engage weaker extensor muscles in subjects with SCI which is crucial in the controlled deceleration required to acquire the target effectively~\cite{Mateo2013}.

It is interesting to note that this study found that individuals with SCI attain higher peak movement velocities compared to healthy controls. The peak velocity for individuals with SCI was found to scale with endpoint target distance from the initial starting point, as shown in Fig. S2. Scaling of velocity with object size and object distance is typical of reach-to-grasp and prehensile grasps and demonstrated by several studies~\cite{Messier1999, Soechting1984}. For individuals with SCI, the velocity scaling factor is significantly higher than for healthy controls demonstrating that the kinematic invariants are preserved, but the ability to control the peak velocity may be altered due to weaker preserved muscles of the forearm~\cite{Bootsma1994}.

\subsection{Limitations of the study}
Our study demonstrates sonomyography-based muscle activity detection in individuals with C5, C6, and C5-C6 cervical spine injuries. Sonomyography utilizes ultrasound to sense the physical muscle deformation in the forearm muscles. Unlike brain-computer interfaces and peripheral neural interfaces, the presence of residual muscle function is critical for muscle activity sensing using sonomyography. Therefore, sonomyography-based techniques for upper-extremity muscle activity sensing in individuals with SCI are limited to those with incomplete motor impairment. 

\section{Conclusion}
Injury to the cervical spinal cord often results in partial loss of motor function in the upper-extremity. Exercise therapy assisted by robotic exoskeletons has been shown to improve motor recovery by augmenting joint movements based on the user’s movement intent. Decoding motor intent to resolve fine levels of motor activity remains challenging due to limitations in non-invasive neural recording technologies. We show that ultrasound imaging of muscle in individuals with cervical spine injury enables dynamic muscle activity sensing. Results demonstrate that with minimal training, users were able to intuitively control an on-screen cursor with ultrasound-derived muscle activity. Ultrasound provides an exciting alternative for the control of bionic interfaces and may further improve the rehabilitation benefits of robotic devices.
\section{References}
\bibliographystyle{IEEEtran}
\bibliography{Bibliography/references}
\end{document}